          [1999/12/02 1.01 (PWD)]
\documentclass[letterpaper,twocolumn]{esapub} % American paper

\usepackage{epsfig}

\newcommand\rsun{R$_\odot$}
\title{What does helioseismology tell us about solar cycle related structural changes 
in the Sun?}
\author{Sarbani Basu}
\affil{Astronomy Department, Yale University, P. O. Box 208101, New Haven CT 06520-8101, U. S. A.}
\affil{email: basu@astro.yale.edu}

\begin{document}


\maketitle

\begin{abstract}
Solar oscillations frequencies show a distinct change with solar activity.
The changes in frequencies can be used to study the time variation of
solar structure.  We discuss constraints on the changes in
solar structure with time as obtained with helioseismic data
covering the last six years.
The frequency variations appear to be dominated by changes in 
the near-surface layers
rather than by changes in the structure of the deeper layers.
\end{abstract}

\section{Introduction}
\label{sec:intro}

The study of the Sun's variability has been of interest to
both geophysicists and astrophysicists.  For geophysicists
the interest lies in studying the effect on earth's climate.
For astrophysicists, the Sun is a fairly typical star with
the advantage of proximity which allows the detailed study of
many phenomena which are important to stellar physics, among
them being magnetic activity.

The most important change that takes place in the Sun on a
time-scale shorter that the human life-span, and hence, on a
time-span that
can directly affect us,  is the solar
magnetic cycle with a periodicity of 11 years.
The total energy output of the Sun has been precisely measured for
about two decades now. These measurements show that solar
irradiance varies with solar cycle, being highest during maximum
activity (Willson \& Hudson 1988). Short-term changes on days to
months are directly related to the evolution of active
regions (Fr\"ohlich \& Pap 1989). 
 In this increasingly technological
age the effect of the solar
magnetic cycle is also being felt as the effect of magnetic storms which
can overload power-lines. Thus one of the aims of studying
solar variability has been to predict solar activity.
There is however, no reliable way to predict solar activity since
there is as yet no accepted model of the solar dynamo.

Till very recently, the only way to study the solar cycle was
to observe changes that take place at the surface and make
models which reproduce the surface phenomena. However, the
history of studying solar rotation has shown us how misleading
just reproducing surface phenomena can be (see, e.g., the review by
Libbrecht \& Morrow 1991).
Helioseismology has given us the means to study what happens
inside the Sun. Solar oscillations frequencies are known to change
with time,  therefore, we now have the ability to
see changes taking place within the Sun even though we
cannot yet predict them.
We can now follow the changes as a function of time and of
both local and global magnetic activity.
Helioseismic data thus give strong constraints on the physical
mechanisms that might be postulated to explain solar variability.

Solar oscillations are thought to be linear and adiabatic (although 
adiabaticity breaks down at the outermost layers, where the
time scale of heat transfer is smaller than the period of the 
oscillations). Different modes of oscillation are
described by three numbers, the radial order $n$ which 
is usually the number of nodes in the radial direction. The
angular eigenfunctions are described by spherical harmonics
$Y^l_m$, where $l$ is the degree and $m$ the azimuthal order,
and $-l\le m \le l$.
The degree $l$ is the
number of nodal lines on the surface, while $m$ is the 
number of nodes along the equator. Details of the
properties of solar oscillations may be found in Christensen-Dalsgaard
(1998).

The most complete information on the oscillations is in principle
contained in the individual frequencies $\omega_{nlm}$
or frequency splittings $\omega_{nlm} - \omega_{nl}$ between modes
in the same multiplet, $\omega_{nl}$ being the average multiplet frequency,
which for cases of slow rotation (as is the case for the  Sun) depends only
on the structure of the Sun.
However, it is more usual to parameterize the rotational splittings in
terms of splitting coefficients according to the formula
\begin{equation}
{\omega_{nlm} \over 2 \pi} \equiv \nu_{nlm}
= \nu_{nl} + \sum_{j=1}^{j_{\rm max}} a_j (n,l) \, {\cal P}_j^{(l)}  \; ,
\label{eq:eq1}
\end{equation}
where the basis functions are polynomials related
to the Clebsch-Gordan coefficients (Ritzwoller \& Lavely 1991, Schou 
et al. 1994).
Thus often instead of speaking of frequency splittings,
one talks of ``$a$ coefficients.'' Because of the
structure of the oscillation eigenfunctions, the values of odd-order $a$ coefficients
are determined  by solar rotation, while those of even-order
coefficients are  determined by   a number of reasons, such as,
deviations from spherical symmetry, large scale flows, magnetic fields, etc.
The information on all the angular dependences of solar structure and dynamics are
in the $a$ coefficients. The mean frequency $\nu_{nl}$ depends only on
the spherically symmetric structure of the Sun.

In this paper we discuss what we have learned about changes that take place 
in the solar interior because of the solar activity cycle. We discuss
pre-SOHO and pre-GONG results in \S~\ref{sec:pre},
the new data used in this work are described in  \S~\ref{sec:new}, inversion
techniques and results are described in \S~\ref{sec:inv}, a brief
description of changes in solar dynamics is given in \S~\ref{sec:dyn},
and we state and discuss our conclusions in \S~\ref{sec:con}.

\section{Pre-SOHO and pre-GONG results}
\label{sec:pre}

That solar oscillations frequencies change with time has been known since the
last solar cycle. Elsworth et al.~(1990) studied low-degree
modes ($l=0,1,2,3$) of the Sun as obtained by the Birmingham 
Solar Oscillations Network (BiSON).
They showed that the frequencies of low degree modes have varied from 1977 to 1988 in a manner 
that is correlated with solar activity as measured by the sunspot number. 
The frequency variation in that period had a peak-to-peak amplitude of
$0.46 \pm 0.106\mu$Hz. They speculated that the change in frequencies
could reflect variations in the solar radius or in the sound speed in the
Sun, and these in turn might be due to changes in solar temperature
and/or magnetic fields. The BiSON results continue to track solar activity.

\begin{table}[htb]
\begin{center}
 \caption{\em Data sets used
\label{tab:tab1}}\vspace{1em}
%    \leavevmode
%\small
\begin{tabular}[h]{lcc}
\hline \\
Set \# & Start Day & 10.7cm Flux$^a$\\
\noalign{\vspace{1em}}
%     &            &  Flux$^a$ \\
%       &           & ($10^{-22}$J/sm$^2$Hz)\\
\hline\\
\multispan{3}{GONG sets\hfill}\\
\noalign{\vspace{1em}}
1--3 & May {\phantom 2}7, 1995 & 75.1\\
4--6 & Aug 23, 1995 & 73.6\\
10--12 & Mar 26, 1996 & 72.7\\
16--18 & Oct 28, 1996 & 73.0\\
24--26 & Aug 12, 1997&  91.0\\
28--30 & Jan {\phantom 2}3, 1998 & 106.2\\
34--36 & Aug {\phantom 2}7, 1998 &  131.3 \\
43--45 & Jun 27, 1999 & 163.4\\
55--57 & Sep {\phantom 2}1, 2000 & 167.3\\
58--60 & Dec 18, 2000 & 166.1\\
\noalign{\vspace{1em}}
\multispan{3}{MDI sets\hfill}\\
\noalign{\vspace{1em}}
1216 & May {\phantom 2}1, 1996 & 72.4\\
1432 & Dec {\phantom 2}3, 1996& 73.2\\
1792 & Nov 28, 1997 & 89.2\\
1936 & Apr 21, 1998 & 108.5\\
2224 & Feb {\phantom 2}2, 1999 & 130.7\\
2440 & Sep {\phantom 2}7, 1999 & 161.1\\
2800 & Sep {\phantom 2}1, 2000 & 175.3 \\
2944 & Jan 23, 2001 & 164.4 \\
3160 & Aug 27, 2001 & 219.8 \\
\hline \\
  \end{tabular}
  \end{center}
\vspace{-1em}
{\it $^a$ Units of $10^{-22}$J s$^{-1}$ m$^{-2}$ Hz$^{-1}$}
\end{table}
\begin{figure}[htb]
  \begin{center}
    \leavevmode
  \centerline{\epsfig{file=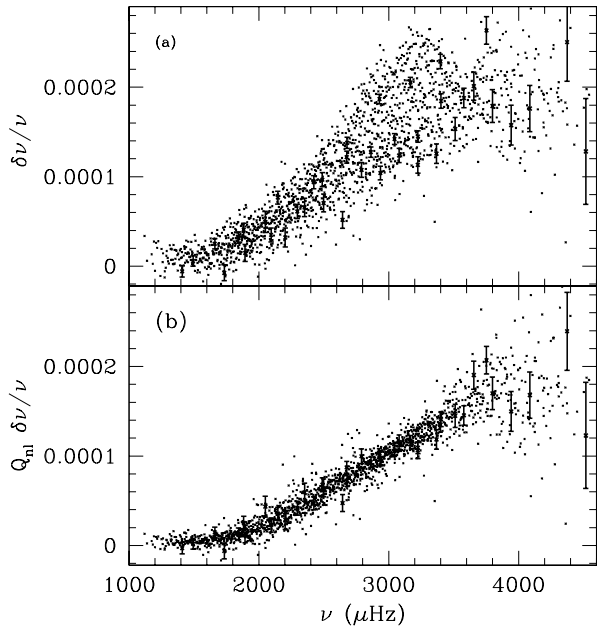,width=8cm}}
  \end{center}
  \caption{\em The frequency differences between MDI sets 3160 and
1216 are shown in Panel (a). Panel (b) shows the same frequency
differences scaled to eliminate the  effects of mode inertia. 
Only a few error bars are shown for the sake of clarity.
}
\label{fig:noav}
\end{figure}

The first results of solar-cycle related changes in intermediate
degree modes
were reported by Libbrecht \& Woodard (1990). 
They showed that measurements of solar oscillation frequencies
in 1986 and 1988 showed systematic differences of the 
order of 1 part in 10,000. The authors 
averaged the frequency differences of modes
with degrees of 5 to 50 at given frequency intervals, after correcting
for differences in mode inertia, and showed
that the averaged frequency difference was just a function
of frequency, and it was a very smooth function of frequency.
This is a strong indication that the cause of the
frequency changes is confined to the near-surface layers of the Sun. 
Any perturbation that is localized at some radius $r_m$ introduces
a perturbation to the frequencies that is proportional to
\begin{equation}
\cos(2\tau_m\omega+\phi),\quad \tau_m=\int^R_{r_m}{dr\over c}
\label{eq:pert}
\end{equation}
(e.g., Gough 1990), where $\tau_m$ is the acoustical depth of the
perturbation, $c$ is the sound speed, $\omega$ the angular frequency 
of the oscillation and $\phi$ a phase. Hence the perturbation
contributes to the frequency differences a component which is an
oscillatory function of $\omega$ with a `frequency' $f=2\tau_m$
which increases with the depth of the perturbation.
Thus provided they have no intrinsic frequency dependence of their 
own, changes in the near surface layers
are expected to introduce frequency shifts that are slowly varying functions
of frequency, which is what is seen in the Libbrecht \& Woodard (1990) data.
The authors also found changes in the even-order $a$ coefficients and
speculated that the changes in frequencies and splittings are a 
result of latitude-dependent changes in the structure of the
sun which are correlated with the Sun's magnetic-activity cycle.

The Mark-I instrument at the Observatorio del Teide 
have yielded low-degree frequencies  for a period of fifteen years between 1984
and 1999. These data too show that the frequencies track solar 
activity (Jimenez-Reyes et al.~2001). These data also show
a hysteresis like behavior in the frequency differences
between the rising and falling phases of Cycle 22 (Jimenez-Reyes et al.~1998). 

Although oscillation frequencies of low degree modes span a long
time period, the Big Bear data of Libbrecht, Woodard \& Kaufman (1990)
and Libbrecht \& Woodard (1990) are the only widely available intermediate
and high degree data set from the Cycle 22. The situation with Cycle 23
is different, with data from both GONG and SOHO, and hence we are now in a
position to look in detail at changes that take place within
the Sun as the solar cycle progresses.

\begin{figure*}[htb]
  \begin{center}
    \leavevmode
  \centerline{\epsfig{file=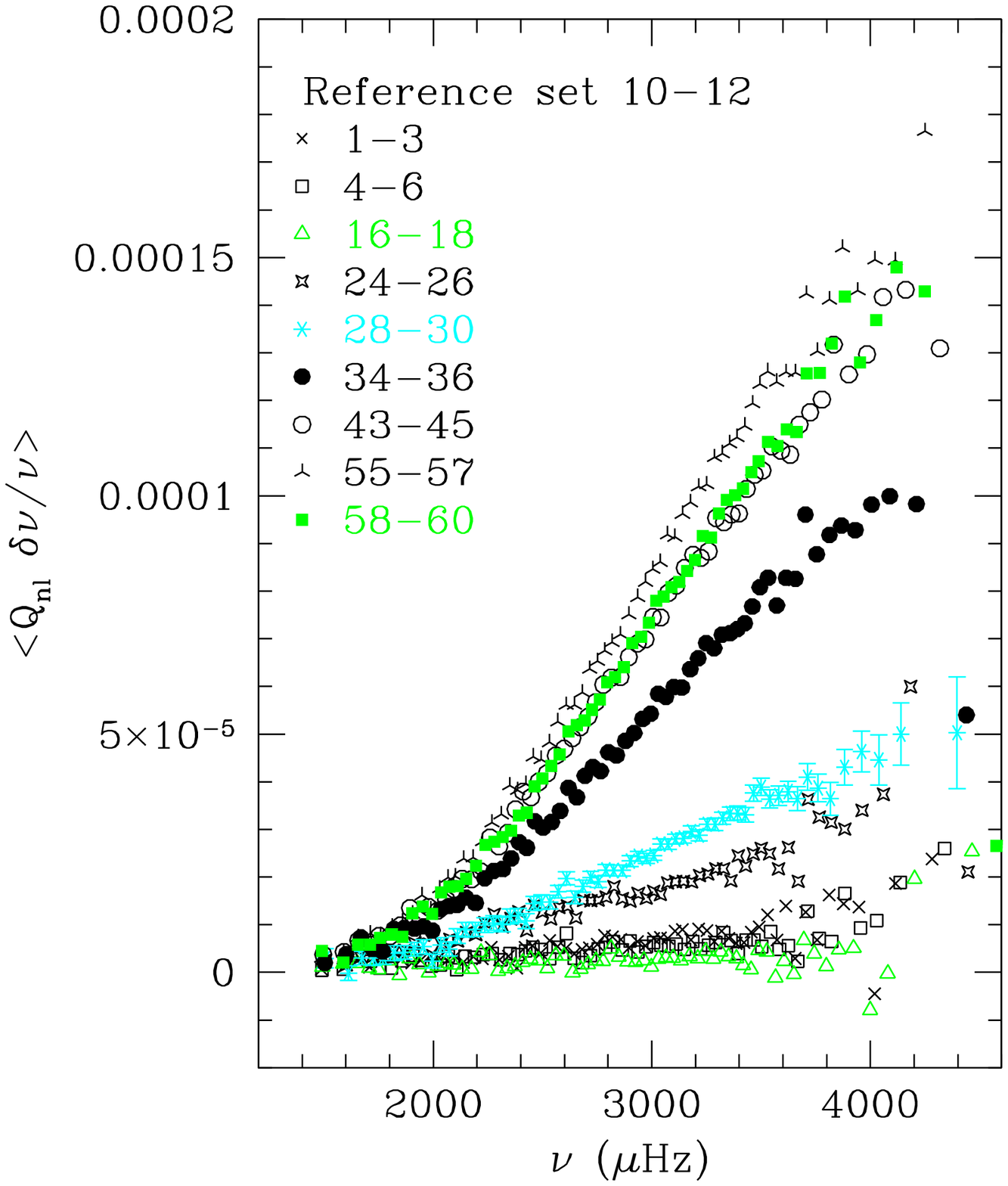,width=8cm}\hfill
\epsfig{file=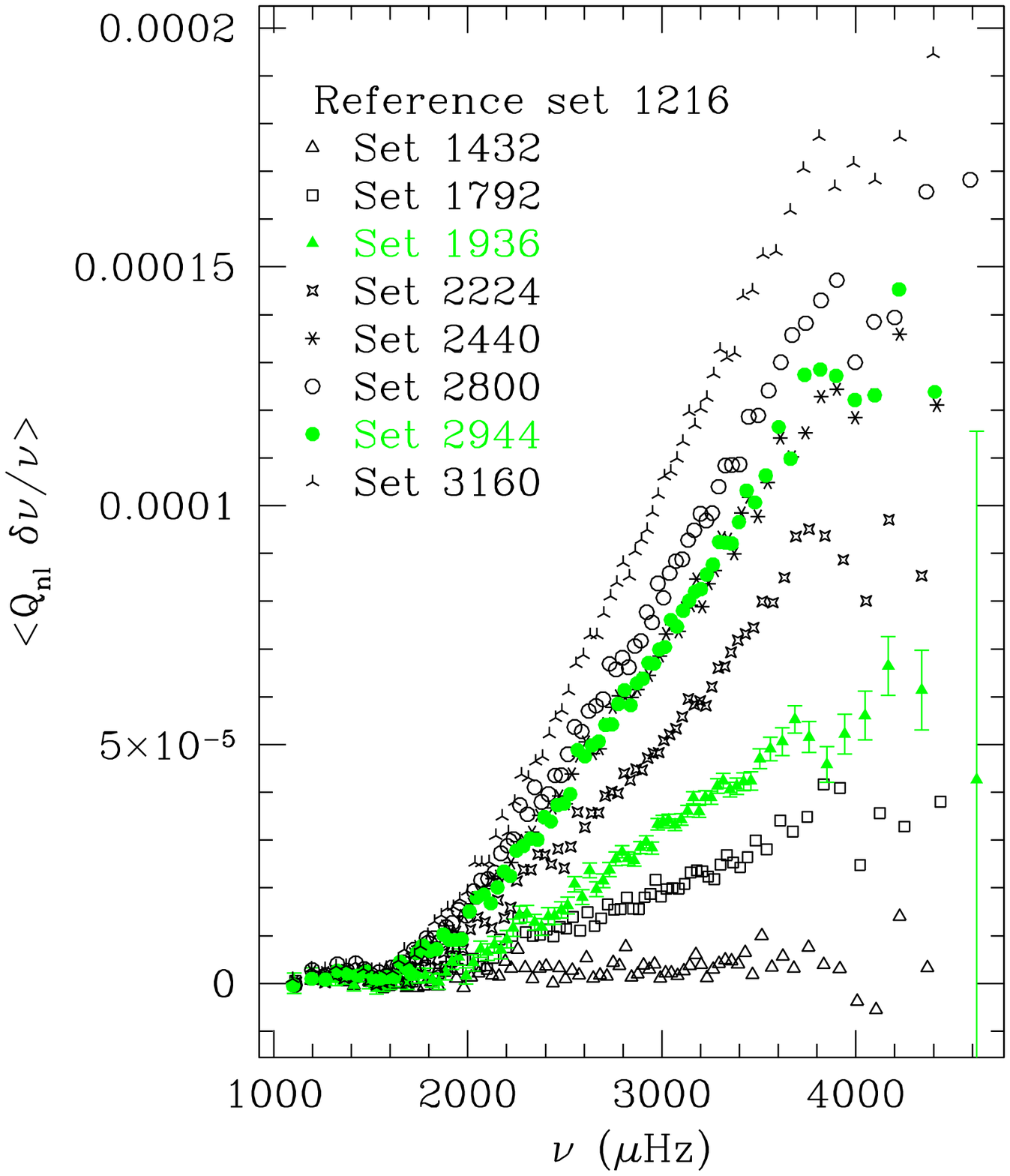,width=8cm}}
  \end{center}
  \caption{\em The averaged, scaled frequency differences between different
different GONG sets and GONG set 10--12 (left panel), and
between different MDI sets and MDI set 1216 (right panel).
}
\label{fig:mdi}
\end{figure*}

\section{New Data}
\label{sec:new}

With the commissioning of the GONG project in 1995 and the helioseismology
instruments on board SOHO, such as MDI, GOLF etc., in 1996 we now are able to study the 
time variation of solar oscillations frequencies in detail. In this work I shall concentrate
on results obtained with GONG and MDI.
The solar oscillations frequencies from the GONG project were obtained from 108 day time series. The 
MDI data were obtained from 72-day time series. Some details
of how GONG and MDI projects obtain the frequencies are available
in Schou et al.~(2002). The data sets used in this work are
listed in Table~\ref{tab:tab1}. The heading ``Start Day'' is the
beginning of the 108 day observational period for the GONG sets
and the 72 day period for the MDI set. Also listed in the table is the average  
10.7 cm radio frequency flux over the observing period. The 10.7 cm radio
frequency flux is known to be a good solar activity index. It should
be noted that only a subset of the available GONG and MDI data
have been used in this work.

Figure~\ref{fig:noav} shows the frequency differences between MDI 
data from late 2001 and mid 1996. Panel (b) of the figure shows
the frequency differences corrected for effect of mode inertia.
Modes with low inertia are easier to perturb than modes with high inertia.
The quantity $Q_{nl}$ is the ratio
of the mode inertia of a mode of degree $l$ order $n$ to that of a mode
of degree 0 with the same frequency as the mode of degree $l$ order
$n$ (Christensen-Dalsgaard \& Berthomieu 1991). Scaling by
$Q_{nl}$ eliminates the mode-mass (hence degree) dependence of the
frequency differences. We see that the scaled frequency differences seem to
be a function only of frequency. To bring out the trends more clearly,
we average the scaled frequency differences in groups of 25. These
averaged frequency differences are shown in 
in Fig.~\ref{fig:mdi} where we show different
GONG and MDI sets. In the GONG figure we have subtracted out the
frequencies of set 10--12 from the other sets. Set 10--12 has the lowest
solar activity level among the GONG sets used. In the MDI figure,
we subtract out the frequencies of set 1216. Again this set has the lowest
activity level.

From Fig.~\ref{fig:mdi} we see that both GONG and MDI data sets
behave in similar ways. We also see that the frequency
differences for the different sets seem to be a fairly smooth function
of frequency. This points to the cause of the differences being fairly
close to the surface of the Sun. We invert the frequency differences
to check for changes in solar structure.

\begin{figure}[t]
  \begin{center}
    \leavevmode
  \centerline{\epsfig{file=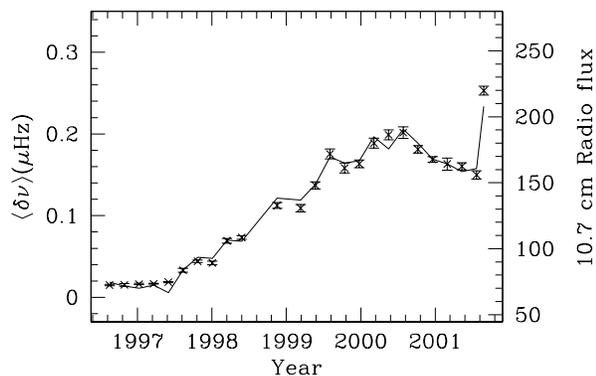,width=8cm}}
  \end{center}
  \caption{\em 
The change in scaled frequencies of different MDI sets plotted 
as a function of time (continuous line). Also plotted is the
10.7 cm radio flux (points). The frequency differences were taken
with respect to set 1216.
}
\label{fig:ac}
\end{figure}

The change in frequency tracks solar activity very closely. In Fig.~\ref{fig:ac}
we show a plot of the average  change in scaled frequency for all MDI modes between 
2 mHz and 3 mHz. The change in frequency was calculated with respect to set 1216.
Also plotted in the figure is the 10.7cm radio flux, and we can see that the
frequency changes and the radio flux, which is an index of solar activity,
track each other very closely. This behaviour is the same as that found
for low degree modes by the BiSON and Tenerife groups (see \S~\ref{sec:pre}).

\section{Inversions for solar structure}
\label{sec:inv}

\subsection{Inversion technique}
\label{subsec:tech}

An inversion for solar structure (e.g. Dziembowski, Pamyatnykh \&
Sienkiewicz 1990; D\"appen et al.~1991; Antia \& Basu 1994;
Dziembowski et al.~1994) generally proceeds through a linearization
of the equations of stellar oscillations around a known reference
model using the variational principle.  The differences  between the
structure of the Sun and the reference model are then related to the
differences in the frequencies of the Sun and the model by kernels.
Non-adiabatic effects and other errors in modeling the surface layers
give rise to frequency shifts (Cox \& Kidman 1984;
 Balmforth 1992) which are not accounted for by the variational
principle.  In the absence of any reliable formulation, these
effects have been taken into account in an {\it ad hoc} manner by
including an arbitrary function of frequency in the variational
formulation (Dziembowski et al.~1990).
Thus the fractional change in frequency of a mode can be expressed
in terms of fractional
changes in the structure of the model and  a ``surface term''.

When the oscillation equation is linearized --- under the assumption
of hydrostatic equilibrium --- the fractional change in the
frequency can be related to the fractional changes in two of the
functions that define the structure of the models.  Thus,
\begin{eqnarray}
{\delta \omega_i \over \omega_i}
& = & \int K_{c^2,\rho}^i(r){ \delta c^2(r) \over c^2(r)}d r +
 \int K_{\rho,c^2}^i(r) {\delta \rho(r)\over \rho(r)} d r\nonumber\\
  & & +{F_{\rm surf}(\omega_i)\over E_i}
\label{eq:inv}
\end{eqnarray}
(cf, Dziembowski et al.~1990).  Here $\delta \omega_i$ is the
difference in the frequency $\omega_i$ of the $i$th mode between the
solar data and a reference model.  The function $c^2(r)$ and $\rho(r)$
are the sound speed and density profiles. 
The kernels $K_{c^2,\rho}^i$
and $K_{\rho,c^2}^i$ are known functions of the reference model which
relate the changes in frequency to the changes in $c^2$ and $\rho$
respectively; and $E_i$ is the inertia of the mode, normalized by the
photospheric amplitude of the displacement.  The term $F_{\rm surf}$,
is the ``surface term'', and 
results from the near-surface errors. The term $F_{\rm surf}$ also
contains information of near surface layers that cannot be resolved
with the data being inverted.

Equation~(\ref{eq:inv}) constitutes the inverse problem that must be solved to infer
the differences in structure between the Sun and the reference model.
The inverse problems can be solved by a number of different
techniques, such as a Regularized Least Squares (RLS) method
(see Antia \& Basu 1994 for an implementation of the method),
or the method of constructing Subtractive Optimally Localized Averages (SOLA)
(see Basu et al. 1996; Rabello-Soares et al. 1999 for implementation). All the results
in this work were obtained with the SOLA method. The different inversion
procedures have been tested thoroughly. In particular, possible errors
introduced in the results due to linearization have been studied
and found to be negligible (Basu et al. 2000).

\begin{figure}[htb]
  \begin{center}
    \leavevmode
  \centerline{\epsfig{file=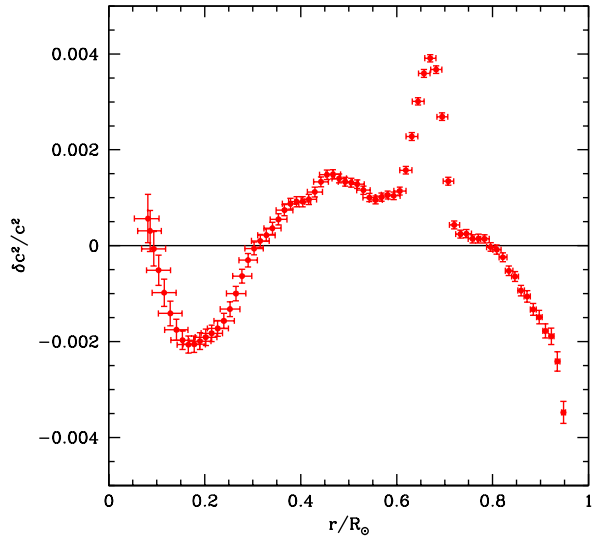,width=8cm}}
  \end{center}
  \caption{\em The differences in squared sound speed between the
Sun and the reference model. 
}
\label{fig:mdi360}
\end{figure}

For all inversions, we have used Model S of
Christensen-Dalsgaard et al.~(1996) as our reference model.
To give a flavour of what the inverted sound-speed difference results
look like, we show in Fig.~\ref{fig:mdi360} the sound-speed differences
between the Sun and the reference model Model S as obtained by inverting
MDI frequencies obtained from the first 360 days of observations. The
figure shows that in most parts of the solar interior the model is very
similar to the real Sun, the sound-speed differences being of the 
order of 0.1\%. However, there is a noticeable difference just below
the base of the convection zone (hence forth CZ, which is at approximately 0.71\rsun).
The solar sound speed is higher in this region. This feature is believed
to be caused by an excess of helium below the base of the CZ
 in the model. Helium diffuses below the  CZ base
because of gravitational settling. Standard solar models do not have
any mixing mechanism below the CZ base hence helium builds up, increasing the
mean molecular weight and hence decreasing the sound speed. The base
of the CZ in the Sun is however a region of large shear, this is the
location of the tachocline. The shear  mixes the material below the
CZ base and hence the decrease in the solar sound speed is less than
the decrease in the sound speed of the model.

\subsection{Inversion results}
\label{subsec:res}

\begin{figure}[htb]
  \begin{center}
    \leavevmode
  \centerline{\epsfig{file=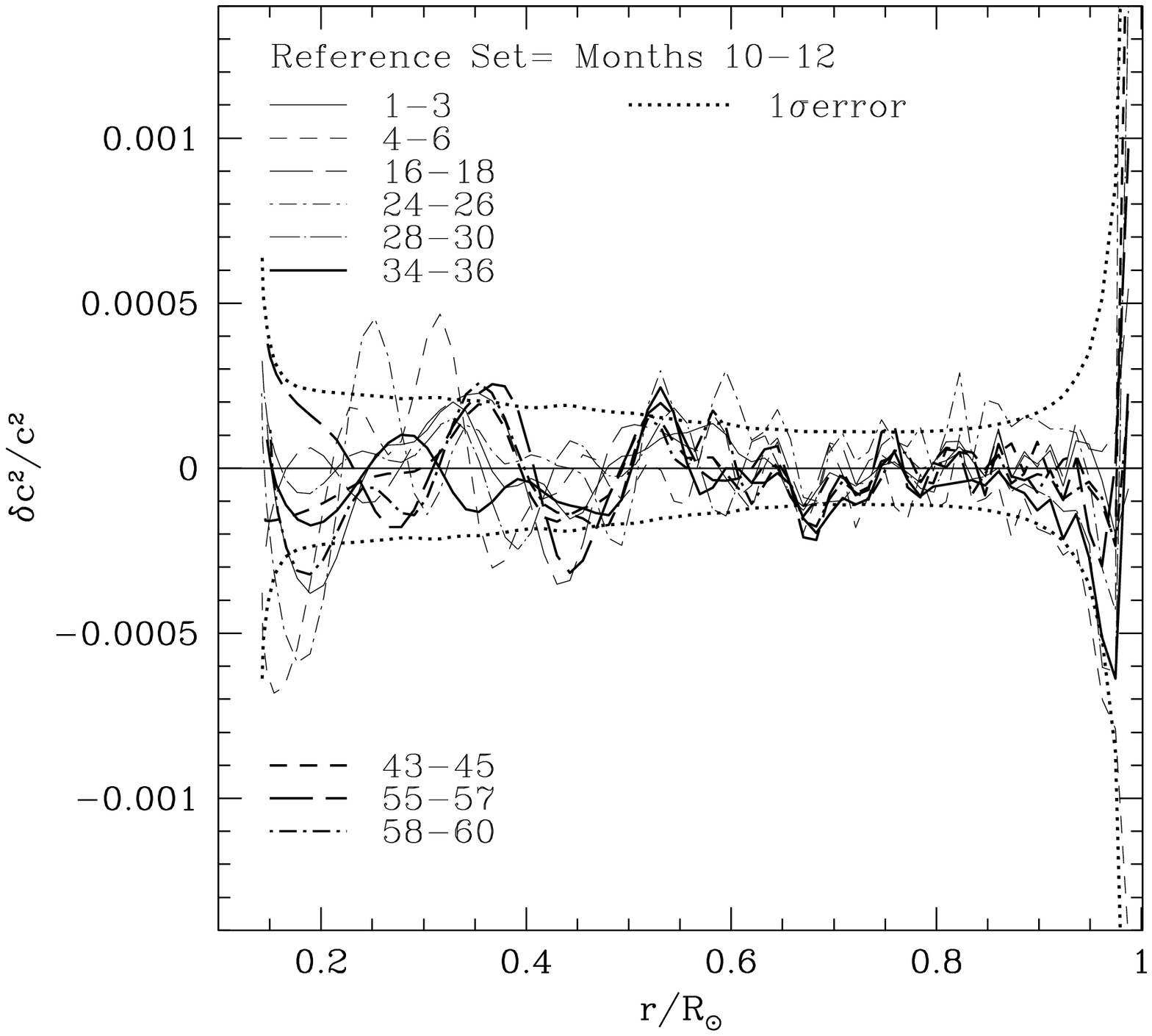,width=8cm}}
  \centerline{\epsfig{file=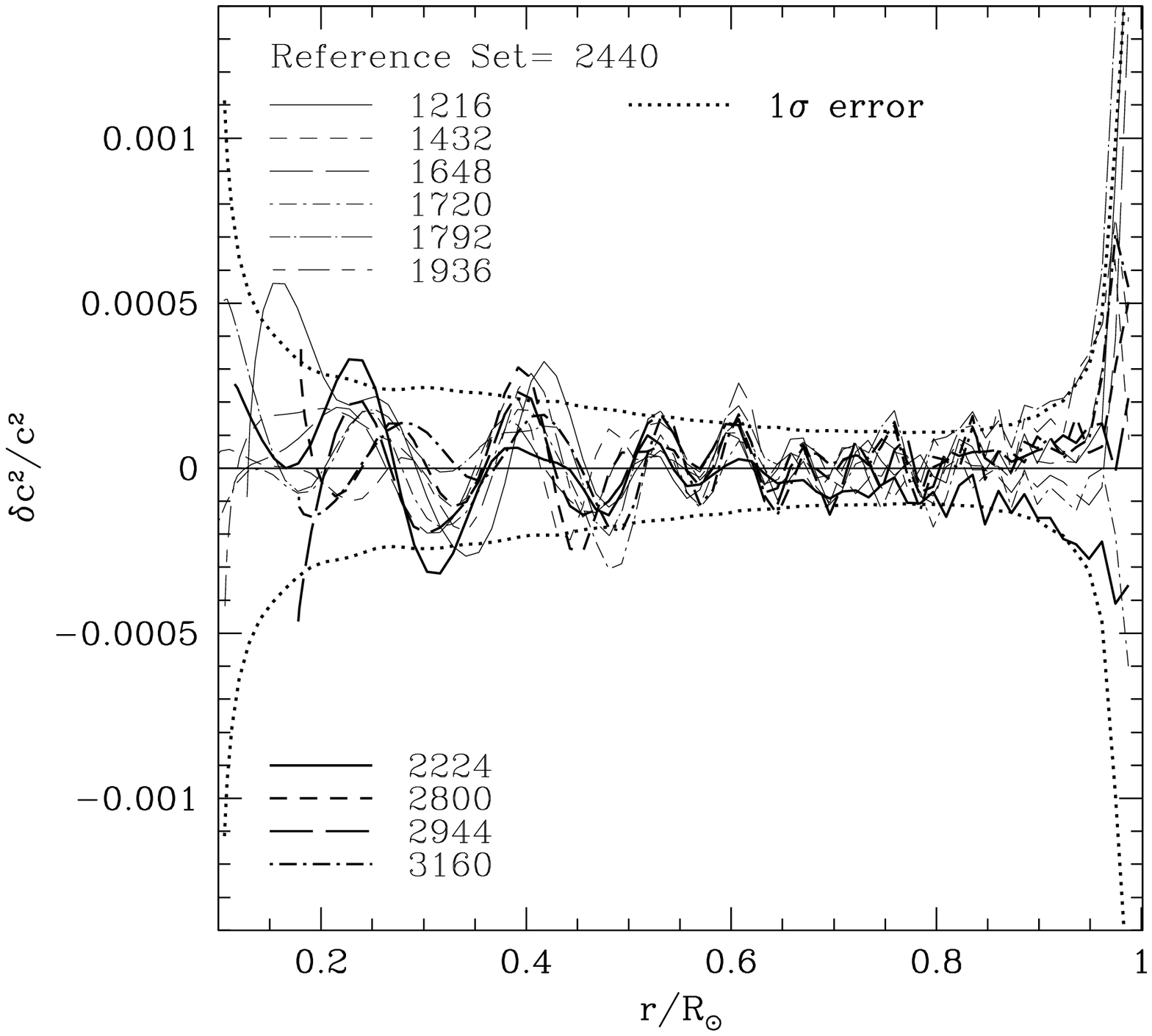,width=8cm}}
  \end{center}
%  \caption{\em The difference in sound-speed obtained by inverting the 
%GONG frequency changes  shown in the left panel of 
%Fig.~\ref{fig:mdi}.
  \caption{\em The difference in sound-speed obtained by inverting the 
GONG frequency changes (upper panel) and MDI
frequency differences (lower panel).
}
\label{fig:csqgong}
\end{figure}

We are mainly interested in seeing how the results shown in Fig.~\ref{fig:mdi360}
change with time. Since the inversion process is linear, instead of inverting
the frequencies of each set and subtracting the inverted results to
find the difference, we invert the frequency differences.
If there are no changes in solar sound speed, the inversion results 
should be zero, within limits of error in the inversion results due
to errors in the frequencies.
There seems to be some systematic differences in the
high degree modes of the pre- and post-recovery MDI sets. Since we
are not yet sure whether the differences are real or are due to errors,
we truncate the MDI mode set above $l=120$ in order to carry out
the inversions. For both GONG and MDI data, we restrict the inverted  mode set 
to modes with frequencies below 3.5 mHz to avoid large surface terms.

%\begin{figure}[]
%  \begin{center}
%    \leavevmode
%  \centerline{\epsfig{file=csqmdi.ps,width=8cm}}
%  \end{center}
%  \caption{\em The difference in sound-speed obtained by inverting the
%MDI frequency changes  shown in the right panel of 
%Fig.~\ref{fig:mdi}.
%}
%\label{fig:csqmdi}
%\end{figure}

We show the sound speed differences obtained by inverting the frequency differences
 in Fig.~\ref{fig:csqgong}. These inversion 
results are reliable from
about 0.1\rsun\ to 0.95\rsun. The radius range is imposed
by the limitations of the mode sets used in the inversions. We cannot get 
reliable results at smaller radii due to the lack of 
sufficient numbers of low degree modes. We cannot get reliable
results for radii larger than 0.95\rsun\ due to lack of high degree modes.
We see from Fig.~\ref{fig:csqgong} that
all the  inversion results seem to lie within the
1$\sigma$ error limits. We do not see any systematic change in the sound speed
differences with time at any radius,  and hence, conclude that at least 
within error limits, there is 
no temporal change in the structure of the Sun.

There is some deviation from the 1$\sigma$ limit in the results of the
outer convection zone, however,  we cannot conclude 
that the changes are due to changes in solar  activity. There are two main reasons
that can cause spurious deviations from 
1$\sigma$ close to the surface. The first reason is that we have not included
high degree modes in the inversions. The MDI sets, as has been discussed earlier
have been restricted to modes with $l \le 120$, the GONG sets are
limited to $l \le 150$. Unfortunately,
it is very difficult to determine the frequencies of very high degree
modes through global analyses due to the systematic errors
involved (e.g., Rhodes et al.~1998). Since inversions at any given radius
is most sensitive to modes with their lower turning points at that radius,
we are unable to invert the outer regions of the Sun reliably. The second reason is
correlation of errors. Even if errors in the input frequencies are uncorrelated,
the inversion process correlated the errors in the inversion results at one
radius with errors in inversion errors at others (see Howe \& Thompson 1996).
This problem   can be severe at large
radii (see Rabello-Soares et al.~1999). This causes the inversion
results at large radii to be pulled on one side of the true solution
instead of being randomly distributed on both sides of the true solution.
The effect is of the order of $1\sigma$ which is seen in the inversion results
in Fig.~\ref{fig:csqgong}.

\subsection{The base of the convection zone}
\label{subsec:cz}

\begin{figure}[t]
  \begin{center}
    \leavevmode
  \centerline{\epsfig{file=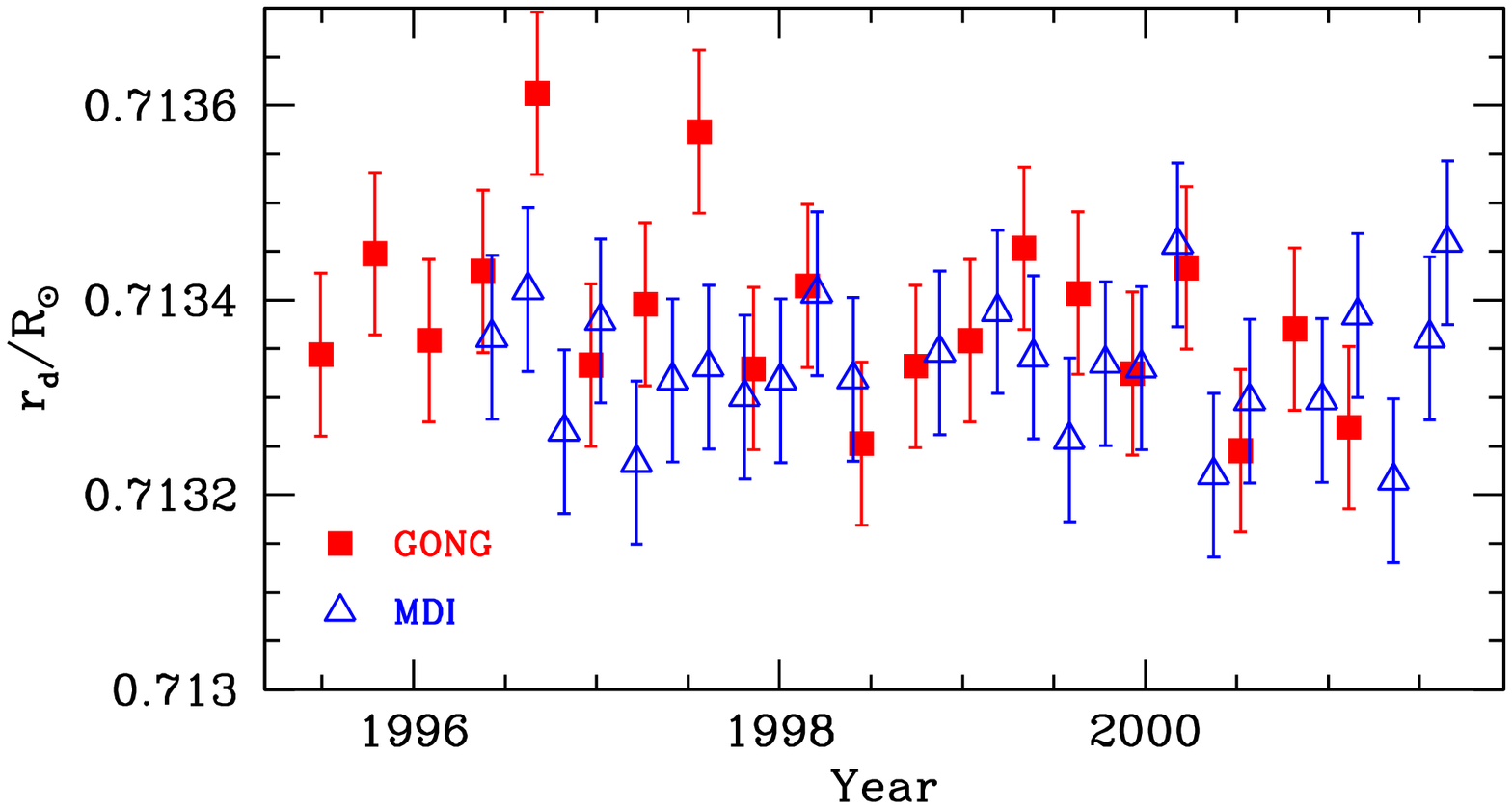,width=8cm}}
  \centerline{\epsfig{file=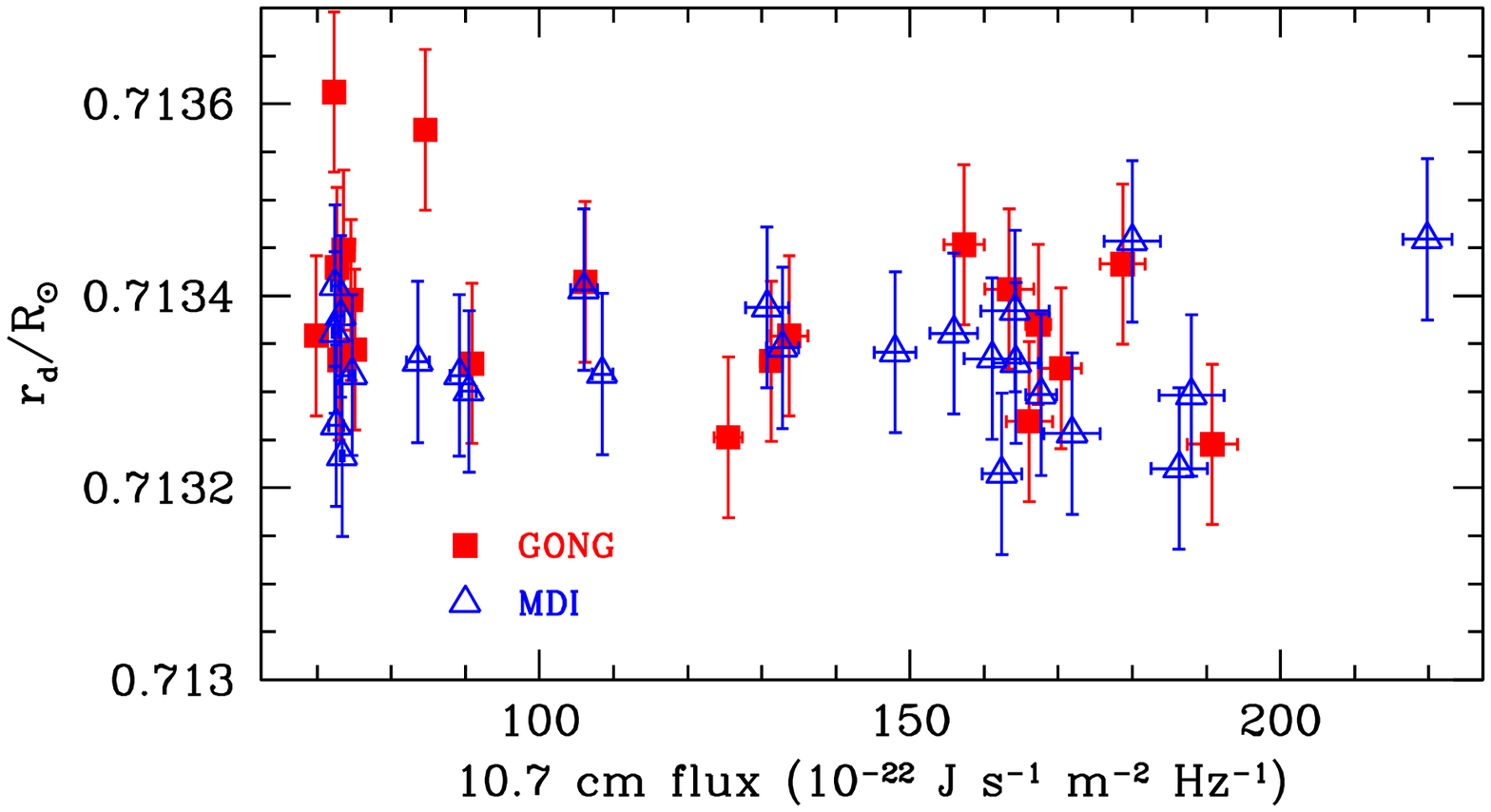,width=8cm}}
  \end{center}
  \caption{\em The position of the base of the solar convection zone as
a function of time and of the 10.7 cm radio flux.
}
\label{fig:cz}
\end{figure}

The seat of the solar dynamo is believed to be close to
the base of the convection zone.
Eff-Darwich \& Korzennik  (2000) in a preliminary analysis of solar cycle
related changes in the Sun reported changes at the base of the convection zone.
Our inversion results in Fig.~\ref{fig:csqgong} fail to show any systematic
change at the CZ base. However, there are better ways to look at the CZ base
more closely.

On seeing the results of Eff-Darwich \& Korzennik  (2000), Basu \& Antia (2001)
investigated the CZ base further, in particular,
to see if there is any change in the position of the base of the CZ,
which can be determined quite precisely from helioseismology
(see Basu \& Antia 1997 for details of the method). 
Basu \& Antia (2001) failed to see any change in the position of the CZ base.
We have extended the work to newer data sets here. The position of the
CZ base as a function of time and solar activity are shown in Fig.~\ref{fig:cz},
and we cannot detect any change in the position of the CZ base either as a 
function of time or as a function of solar activity. 
Basu \& Antia (2001) did not find any changes in the extent of overshoot
below the CZ base either. The lack of change at the base of the CZ is
not altogether surprising. If we assume that the changes are caused by magnetic
fields, very high magnetic fields  to change the structure
by adding to the gas pressure of that layer.

\subsection{The latitudinal distribution of sound speed}
\label{sumsec:asph}

\begin{figure}[htb]
  \begin{center}
    \leavevmode
  \centerline{\epsfig{file=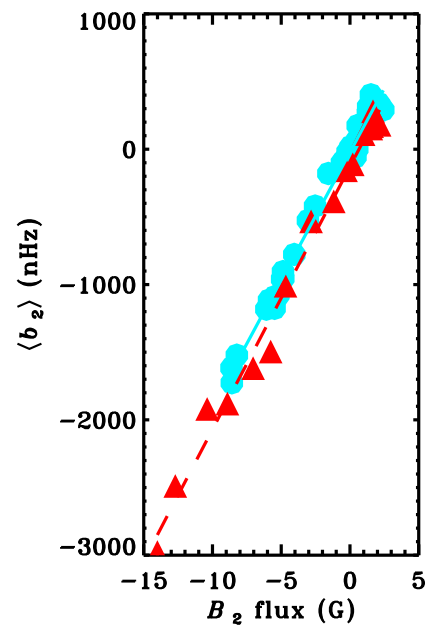,width=6.5cm}}
  \end{center}
  \caption{\em The average $a_2$ coefficients of different data sets plotted as
a function of the corresponding Legendre component of the solar  surface magnetic
field. Both GONG (triangles) and MDI(circles) data are shown.
}
\label{fig:acoef}
\end{figure}
\begin{figure}[htb]
  \begin{center}
    \leavevmode
  \centerline{\epsfig{file=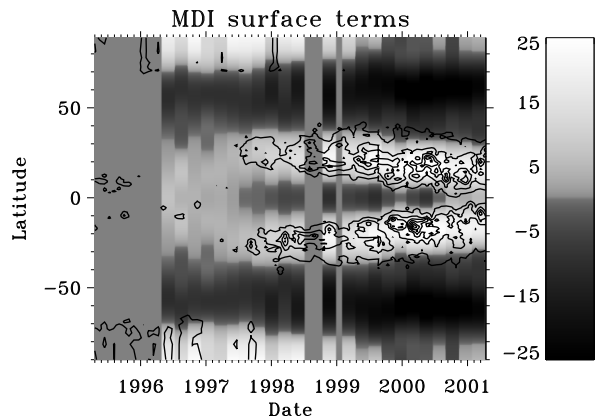,width=8cm}}
  \end{center}
  \caption{\em 
Grey scale map showing the reconstruction of the latitudinal dependence
of the surface term from  MDI, multiplied by 100. Overlaid contours show the
Kitt Peak unsigned magnetic flux with the $B_0$ term subtracted; contour spacing is 10G
}
\label{fig:sur}
\end{figure}

Howe, Komm and Hill (1999)
have found linear relations between
the even-order $a$ coefficients and the Legendre decomposition components
of the surface magnetic flux up to $a_8$,
and showed that these relations could be extended backwards in time
to the BBSO data from the previous cycle. This can be see in Fig.~\ref{fig:acoef}
for the $a_2$  coefficient. The coefficients have been averaged
as per Eq.~(5) of Antia et al.~(2001), which takes into account the
effect of mode inertia. Antia et al.~(2001) that a similar relation holds
true for all even-order coefficients upto $a_{14}$. Coefficients of
orders higher than 14 were not studies.

Inversions for the latitudinal distribution of sound speed show no 
changes with time (Antia et al.~2001). This leads to the question again whether
all the changes in the even $a$ coefficients are confined to changes in the
surface term. This indeed seems to be the case. A plot of the surface term
obtained from the inversions is shown in Fig.~\ref{fig:sur}. Superposed
on the plot is the surface magnetic field and we can see a clear correlation
between the two. Thus as in the case of the frequencies, we are forced to
conclude that changes in the even-order $a$ coefficients are caused by 
changes in the near-surface regions of the Sun.

\section{Changes in solar dynamics}
\label{sec:dyn}

Although all changes in solar structure seems to be confined to the outer layers
which we are as yet unable to resolve in inversions, changes in solar
dynamics can be detected fairly easily. 

Solar zonal flows show a change in with time (see Schou 1999, Howe et al. 2000a,
Antia \& Basu 2000, Antia \& Basu 2001, Basu \& Antia, this volume).
The different low- and intermediate-latitude bands migrate towards the
equator, while high latitude bands migrate towards the poles 
(Antia \& Basu 2001, Basu \& Antia, this volume). Antia \& Basu (2000)
found that these flows penetrate to about 0.1\rsun, similar results are
found by Howe et al. (2000a).

Results of changes deeper in the Sun are more controversial. Howe et al.(2000b)
claim to see a 1.3 year oscillation in the solar rotation rate at the 
base of the CZ, which Antia \& Basu (2000), Corbard et al (2001) 
and Basu \& Antia (this volume) fail to find.

Solar meridional flows change too. The maximum speed of the flow shows a
decrease with increase in activity (Basu \& Antia, this volume). In addition
there are more complex changes on short time-scales (Haber et al. 2001).

\section{CONCLUSIONS}
\label{sec:con}

Newer more precise solar oscillation frequencies have confirmed that
these frequencies change with time. The frequencies show an
increase with solar activity and changes are  found to be
remarkably well correlated with solar activity indices.  The
change in frequency is almost purely a function of the
frequency of the mode, once mode inertia effects are taken into account.
This points to changes confined to the near-surface layers of the
Sun as being the cause of the observed changes in oscillation
frequencies.
We have only looked at low and intermediate
degree modes in this work. High degree modes obtained from
local area analysis (See Bogart et al. this volume) show that
changes in the frequencies of high degree modes in active regions
are not merely functions of frequency, indicating the possibility
of structural changes in the solar interior, in particular
in the near-surface regions that are not resolved properly
in current inversions.
Frequency splittings also change with activity. In particular, the even
order splittings are almost linearly correlated with the 
corresponding component of the solar magnetic field. 

Results obtained by inverting the  frequencies obtained at different epochs
show no significant changes in the structure of the solar interior, at least
up to  a radius of $0.95$\rsun. 
Data sets used in this work  do not contain
enough high degree modes to allow a reliable inversion of shallower
layers. All changes are consistent as being due to
errors in the frequencies. We do not see any changes at the base of the
solar convection zone.

The latitudinal distribution of solar sound speed does not show
any changes with time. The ``surface term'' is however, highly correlated
with the latitudinal distribution of the solar surface magnetic field. This
result seems to suggest that the frequencies and splitting coefficients are changed
by the direct effects of magnetic fields.

Although solar structure does not appear to
show changes with time and activity, solar flows do. Both zonal and meridional flows
show changes with time.

\section*{ACKNOWLEDGMENTS}

This work  utilizes data obtained by the Global Oscillation
Network Group (GONG) project, managed by the National Solar Observatory
which is
operated by AURA, Inc. under a cooperative agreement with the
National Science Foundation. The data were acquired by instruments
operated by the Big Bear Solar Observatory, High Altitude Observatory,
Learmonth Solar Observatory, Udaipur Solar Observatory, Instituto de
Astrofisico de Canarias, and Cerro Tololo Inter-American Observatory.
This work also utilizes data from the Solar Oscillations
Investigation/ Michelson Doppler Imager (SOI/MDI) on the Solar
and Heliospheric Observatory (SOHO).  SOHO is a project of
international cooperation between ESA and NASA.
MDI is supported by NASA grants NAG5-8878 and NAG5-10483
to Stanford University.

\end{document}